# Solving the paradox of the solar sodium D$_1$ line polarization

Ernest Alsina Ballester,[1, 2, *] Luca Belluzzi,[1, 3, 4] and Javier Trujillo Bueno[2, 5, 6]

[1]*Istituto Ricerche Solari (IRSOL), Università della Svizzera italiana, 6605 Locarno-Monti, Switzerland*
[2]*Instituto de Astrofísica de Canarias (IAC), 38205 La Laguna, Tenerife, Spain*
[3]*Leibniz-Institut für Sonnenphysik (KIS), 79104 Freiburg, Germany*
[4]*Euler Institute, Università della Svizzera italiana, 6900 Lugano, Switzerland*
[5]*Departamento de Astrofísica, Universidad de La Laguna, 38206 La Laguna, Tenerife, Spain*
[6]*Consejo Superior de Investigaciones Científicas (CSIC), Spain*
(Dated: August 31, 2021)

Twenty-five years ago, enigmatic linear polarization signals were discovered in the core of the sodium D$_1$ line. The only explanation that could be found implied that the solar chromosphere is practically unmagnetized, in contradiction with other evidences. This opened a paradox that has challenged physicists for many years. Here we present its solution, demonstrating that these polarization signals can be properly explained in the presence of magnetic fields in the gauss range. This result opens a novel diagnostic window for exploring the elusive magnetism of the solar chromosphere.

Observations of quiet regions of the solar disk (i.e., outside areas of strong magnetic activity) with high-sensitivity spectropolarimeters reveal that the entire solar spectrum is linearly polarized, especially close to the edge of the Sun's visible disk [1–3]. The physical mechanism responsible for this linear polarization is the scattering of anisotropic radiation within the solar atmosphere. In spectral lines, this so-called scattering polarization is due to the presence of atomic level polarization (i.e., population imbalances and quantum interference between the magnetic sublevels of the atomic energy levels), produced when the atom is illuminated by anisotropic radiation (i.e., anisotropic optical pumping) [4, 5]. Atomic level polarization is sensitive to collisions with neutral hydrogen atoms [6] and to the presence of magnetic fields through the Hanle effect [7, 8]. These mechanisms are especially efficient in relaxing the polarization of long-lived atomic levels, such as the ground states.

Twenty-five years ago, unexpected scattering polarization signals were discovered in the core of the Na I D$_1$ line [1, 2, 9–12], a line transition that was thought to be intrinsically unpolarizable [1, 2]. These enigmatic signals could only be explained by taking the hyperfine structure (HFS) of sodium into account and assuming that the lower level of D$_1$ (the ground state of sodium) has a substantial amount of atomic polarization [13]. Because long-lived atomic levels are particularly vulnerable to the Hanle effect, the required amount of ground-level polarization is incompatible with the presence of inclined magnetic fields stronger than about 0.01 G in the lower solar chromosphere [13], where the core of the D$_1$ line originates [14]. The requirement that the lower solar chromosphere must be practically unmagnetized conflicts with the results from observations in other spectral lines as well as with plasma physics arguments, which instead indicate the presence of magnetic fields in the gauss range in this key interface region of the solar atmosphere [15].

The proposed explanation for the unexpected scattering polarization signal observed in the core of the Na I D$_1$ line gave rise to a serious and intriguing paradox in solar physics, which has remained unresolved since its introduction in 1998 [13]. In that original investigation, the anisotropic radiation field that pumps the atoms of the solar atmosphere was assumed constant with wavelength over the very small spectral interval spanned by the nearby hyperfine structure (HFS) components of both the sodium D$_1$ and D$_2$ lines. When this apparently reasonable assumption is made (see Fig. 1), the four HFS transitions of the D$_1$ line are all pumped by the same D$_1$ line radiation field. Under this hypothesis, it is found that it is impossible to induce atomic polarization in the upper HFS F-levels of the D$_1$ line, unless atomic polarization is also present in the F-levels of the ground state of sodium [13]. In this scenario, the upper F-levels of the D$_1$ line are consequently depolarized by the same very weak magnetic fields that depolarize the F-levels of the ground state [16].

Solving the paradox of the solar sodium D$_1$ line is thus a very important step forward in our understanding of the physical processes that produce polarization in spectral lines. This is essential for deciphering the magnetism of the quiet solar chromosphere through the modeling of the unprecedented spectropolarimetric observations that the new generation of solar telescopes, such as the upcoming National Science Foundation's Daniel K. Inouye Solar Telescope [17], will soon provide. As a matter of fact, over the last two decades the solar physics community has witnessed an intense research activity motivated by the enigmatic scattering polarization of the solar Na I D$_1$ line. This paradox even led to questioning the quantum theory of scattering, and motivated optical pumping laboratory experiments using monochromatic (laser [18]) or spectrally flat (halogen bulb [19]) incident light, as well as theoretical investigations in which the atomic system is assumed to be excited by monochromatic [20, 21] or broadband [16] radiation.

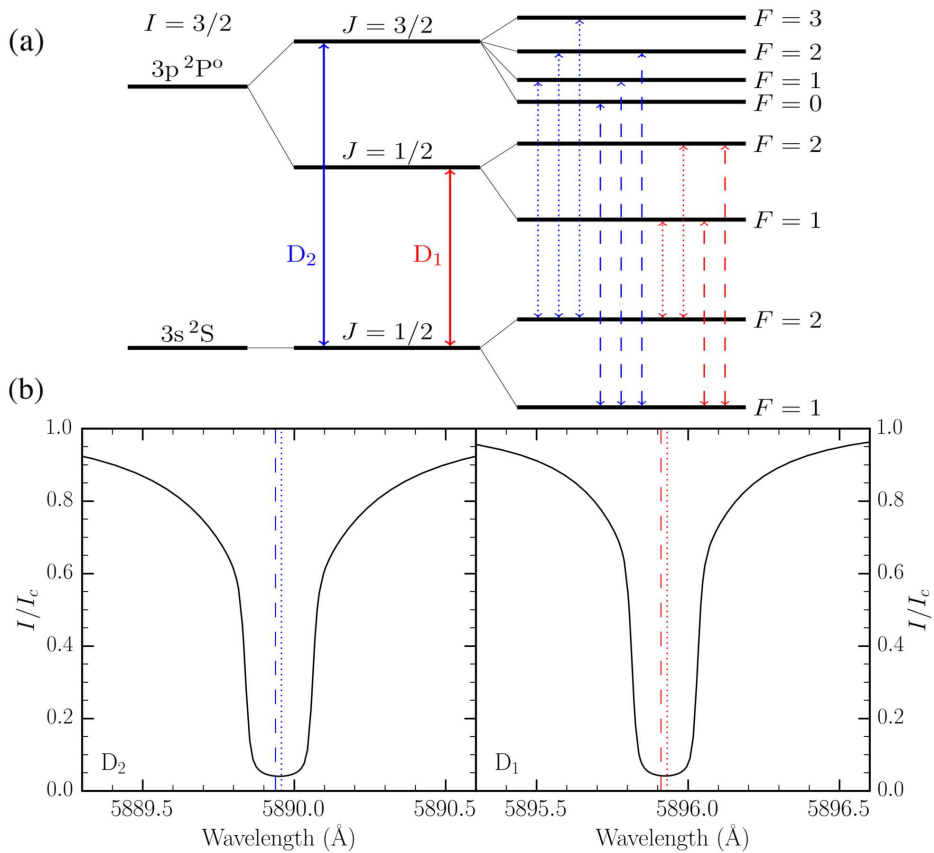

FIG. 1. (a) Grotrian diagram of the considered atomic model of sodium. The diagram shows the upper and lower terms of the sodium doublet, as well as the fine structure and hyperfine structure (HFS) levels (energy splittings not to scale). The fine structure transitions corresponding to the $D_1$ and $D_2$ lines are also indicated, as well as the allowed transitions between HFS $F$-levels. The quantum number $I$ indicates the nuclear spin. (b) Theoretical intensity profiles (normalized to the continuum) of the $D_2$ and $D_1$ lines. Calculations were carried out in a one-dimensional semiempirical model of the solar atmosphere [22] for a line of sight with $\mu = \cos\theta = 0.1$ (with $\theta$ the heliocentric angle). The dashed and dotted vertical lines show the wavelength positions of the various HFS components of the $D_1$ and $D_2$ lines, indicated in the Grotrian diagram (a).

In the quest to resolve this challenging problem, it was of particular interest to theoretically identify a mechanism capable of introducing linear polarization in the Na I $D_1$ line without the need of ground-level polarization [23]. This mechanism was identified by taking into account the small variations of the pumping radiation field, and in particular its anisotropy, across the very narrow spectral interval spanned by the various HFS components of the $D_1$ line (i.e., by relaxing the apparently reasonable assumption that the radiation field is constant with wavelength across this narrow interval). However, the employed theoretical formulation was only applicable in a very idealized situation, namely in the absence of elastic collisions and magnetic fields, which could potentially depolarize the $D_1$ line signal [23, 24].

Here we model the polarization of the solar sodium doublet radiation through a rigorous theoretical framework for the generation and transfer of polarized radiation, suitable for taking into account the detailed spectral structure of the radiation field pumping the atoms. This quantum theory of atom-photon interactions [25] allows considering correlations between the incoming and outgoing photons in the scattering events (i.e., partial frequency redistribution phenomena), in the presence of collisions and magnetic fields. By applying this theory, for the first time we could take into account the detailed spectral structure of the radiation, together with the effects of magnetic fields of arbitrary strength and elastic collisions in a realistic atomic model including HFS. We have calculated the intensity and polarization profiles of the radiation emerging from semiempirical models of the solar atmosphere by numerically solving this complex non-equilibrium radiative transfer problem. Details of such calculations can be found in the Supplemental Material [26]. Our results show that linear polarization is produced in the $D_1$ line in the absence of any ground-state polarization, even in the presence of inclined magnetic fields in the gauss range, and that the calculated spectral line polarization is similar to that found in re-cent high-precision spectropolarimetric observations [12].

Figure 2 shows a comparison between the calculated fractional linear polarization and observations taken with the Zurich Imaging Polarimeter (ZIMPOL-3) [27] in a quiet region close to the edge of the solar disk. The Q/I pattern calculated with (blue solid curve) and without (red dotted curve) magnetic fields in a semiempirical model [22] of the quiet Sun is compared with two different measurements (black curves), one covering a spectral range that includes both lines of the sodium doublet (upper panel) and another with a significantly better spectral resolution but for a range containing only the $D_1$ line (lower panel). In both cases, a very good agreement between observations and synthetic profiles is found when including a volume-filling tangled magnetic field of 15 G. For more details on the considered magnetic field, see the Supplemental Material [26].

The physical mechanisms considered in the present Letter produce a conspicuous linear polarization signal in the core region of the sodium $D_1$ line, which survives in the presence of inclined magnetic fields with strengths in the gauss range, in contrast to the linear polarization signals produced by ground-level polarization [13, 16]. Indeed, the agreement with the observations improves when including such magnetic fields in the atmospheric model. In the presence of such magnetic fields, we have verified that not only is a clear line-core polarization signal found for lines of sight close to the limb, but it also remains appreciable closer to the center of the solar disk, in agreement with observational results [9]. For more details, see the Supplemental Material [26]. Through calculations carried out in the absence of magnetic fields, we have also verified that the depolarizing effect of elastic collisions that induce transitions between magnetic sublevels of the same F-level does not have an appreciable impact on the linear polarization pattern in either line of the doublet, even when purposely overestimating the collisional depolarization rates. This is discussed in further detail in the Supplemental Material [26].

As shown in Fig. 2, when magnetic fields with strengths on the order of gauss are present at chromospheric heights, the Stokes $Q/I$ amplitudes in the core of both the $D_1$ and $D_2$ lines decrease appreciably through the action of the Hanle effect, whose impact is significantly greater in the $D_2$ line. Moreover, the scattering polarization signals in the wings of both lines, and especially the local minima found just outside the Doppler core of the $D_2$ line, are sensitive to the presence of magnetic fields in the underlying photosphere through the same magneto-optical effects that introduce a magnetic sensitivity in the scattering polarization wings of stronger resonance lines [28–30]. Interestingly, such magneto-optical effects operate in the wings of the sodium D-lines at field strengths similar to those that produce the Hanle effect in the line core.

We point out that our radiative transfer modeling has been carried out for a one-dimensional semiempirical static model of the solar atmosphere, in which only an inclined magnetic field can break the axial symmetry of the pumping radiation field. In the real solar atmosphere, the spatial gradients of the macroscopic velocity and the horizontal thermal and density inhomogeneities of the plasma can break the axial symmetry without the need of an inclined magnetic field [31], and a larger variety of $Q/I$ shapes is to be expected. Moreover, it is important to note that in dynamical models of the solar chromosphere [32] the anisotropy of the D-line radiation shows a wide range of variation, which may result in $Q/I$ signals of much larger amplitudes [33].

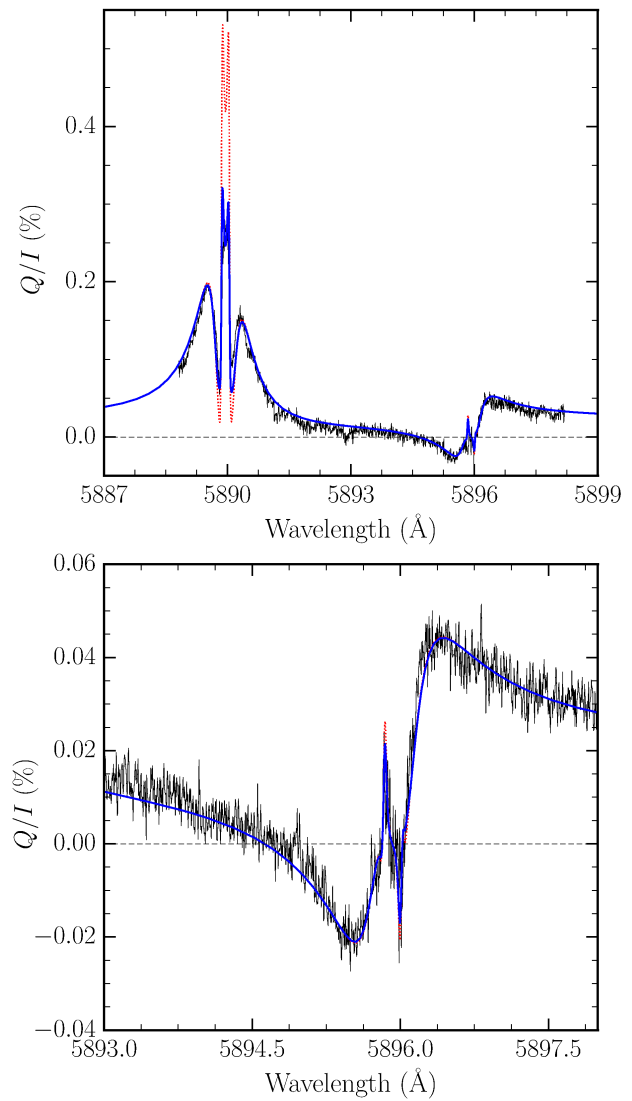

FIG. 2. Fractional linear polarization (Q/I) pattern of the Na I doublet. The black solid curves represent spectropolarimetric observations performed using ZIMPOL-3 [27] in quiet regions close to the solar limb. The measurement shown in the upper panel was taken for a spectral interval containing both $D_1$ and $D_2$, with a spectral sampling of 8.38 mÅ/pixel. The one shown in the lower panel included only $D_1$, with a spectral sampling of 4.60 mÅ/pixel. In both panels, the red dotted curves and the blue solid curves show the results of radiative transfer calculations in a semiempirical one-dimensional atmospheric model of the quiet solar atmosphere [22]. The red-colored profiles were obtained in the absence of a magnetic field, whereas the blue-colored ones were obtained including a volume-filling tangled magnetic field of 15 G throughout the entire atmosphere (see the Supplemental Material [26]). The theoretical Stokes profiles were calculated for a line of sight with $\mu = 0.1$ for the top panel, and $\mu = 0.125$ for the bottom one, in accordance with the observation to which they are compared. The reference direction for positive Stokes Q is taken parallel to the nearest solar limb.

Despite the idealization that a one-dimensional static atmospheric model represents, observed antisymmetric linear polarization signals in the core of the $D_1$ line can be reproduced remarkably well when accounting for the HFS of sodium and for the spectral structure of the pumping radiation field in the presence of magnetic fields. A good agreement with recent observations of high polarimetric accuracy is found when considering isotropic or canopy-like magnetic fields with strengths in the gauss range, as inferred from other chromospheric diagnostics [15]. This provides a satisfactory resolution to the two-decades-long solar sodium paradox and opens up a new window for probing the elusive magnetic fields of the solar chromosphere in the present new era of large-aperture solar telescopes.

This work was supported by the Swiss National Science Foundation (SNSF) through Grant 200021–175997 and by the European Research Council through the European Union's Horizon 2020 research and innovation programme (ERC Advanced Grant agreement No. 742265).

# Supplemental Material to [1]

The Supplemental Material contains eight sections. In Sect. I, details of the considered atomic model are provided. Section II concerns the theoretical framework on which the considered radiative transfer (RT) calculations are based. The redistribution matrix formalism and the assumptions that have been made within it are also discussed. Section III contains a discussion on the treatment of collisions with neutral perturbers and their influence on the spectral line polarization. In Sect. IV, further details on the line and continuum contributions to the RT coefficients are presented. In Sect. V, the iterative scheme used for the RT calculations is discussed. Section VI briefly provides details on the considered atmospheric model. In Sect. VII, various treatments of tangled magnetic fields, including the one made in [1] and in the present Material, are explained. In the same section, the center-to-limb variation of the linear polarization pattern of the Na I D-lines, obtained in the presence of such magnetic fields, is shown. Finally, Sect. VIII presents and discusses a series of RT calculations carried out considering deterministic magnetic fields of various strengths.

## I. THE ATOMIC MODEL

Our RT modeling of the polarization produced by anisotropic radiation pumping in the sodium D-lines, taking into account the full impact of magnetic fields via the Hanle, Zeeman, and magneto-optical (MO) effects, was carried out considering a two-term atom with hyperfine structure (HFS). The upper term ($^2$P$^{\rm o}$) consists of two fine structure $J$ levels, having total angular momentum $J = 1/2$ (upper level of D$_1$) and $J = 3/2$ (upper level of D$_2$). The lower term ($^2$S) consists of the ground level of sodium ($J = 1/2$), which is the lower level of both D$_1$ and D$_2$. This level is treated as unpolarized and all its sublevels are considered infinitely sharp. Due to the interaction with the nuclear spin ($I = 3/2$), the $J$ levels split into various HFS $F$ levels, as shown in Fig. 1 of [1]. We account for quantum interference between pairs of HFS magnetic sublevels (a) of the same $F$ level, (b) of different $F$ levels pertaining to the same $J$ level ($F$-state interference), and (c) of different $F$ levels pertaining to different $J$ levels of the same term ($J$-state interference). Interference (b) produces the depolarization due to the HFS [2] and interference (c) is responsible for the sign reversal of the scattering polarization pattern [3, 4] between the D$_1$ and D$_2$ line centers that can be appreciated in Fig. 2 of [1]. In this paper, we refer to the magnetically induced modification of both interference (a) and (b) as the Hanle effect, rather than restricting its definition only to the impact on interference (a), as is done elsewhere [2, 5]. Because of the large energy separation between the upper $J$ levels of D$_1$ and D$_2$, the impact of the magnetic field on interference (c) is negligible. The splitting of the various HFS magnetic sublevels was calculated in the general incomplete Paschen-Back (IPB) effect regime [2, 6].

We used experimental data for the energies of the upper $J$ levels [7]. The energy splittings of the HFS $F$ levels were calculated using recent experimental values for the HFS constants for the various $J$ levels [8]. For the Einstein coefficient for spontaneous emission from the upper to the lower term, we used $A = 6.15 \cdot 10^7$ s$^{-1}$, obtained by averaging the experimental values for the individual D$_1$ and D$_2$ transitions [7].

## II. THE THEORETICAL FRAMEWORK

We applied a rigorous quantum theory of spectral line polarization [9], which allows us to model the polarization of resonance lines due to the combined action of anisotropic radiation pumping and the Hanle, Zeeman, and MO effects in a two-term model atom with HFS, accounting for frequency correlations between the incoming and outgoing photons in the scattering events (partial frequency redistribution effects). This quantum theory is based on a high-order expansion of the atom-photon interactions, and includes the impact of elastic collisions, which are responsible for both frequency redistribution and changes in the radiatively induced atomic level polarization. We formulated the equations of the problem in terms of the redistribution matrix [9, 10], which we transformed from the atomic rest frame into the observer's frame taking into account the Doppler redistribution due to small-scale atomic motions, assuming a Maxwellian distribution of velocities. In this reference frame, we adopted the angle-averaged assumption [11] for the coherent part ($R_{\rm II}$) of the redistribution matrix, and we considered the limit of complete frequency redistribution in the observer's frame [12] for the incoherent part ($R_{\rm III}$). These are suitable choices within the scope of this investigation.

## III. THE IMPACT OF COLLISIONS WITH NEUTRAL PERTURBERS ON ATOMIC POLARIZATION

Collisions with neutral perturbers (mainly hydrogen atoms in the solar atmosphere) can induce transitions between the magnetic sublevels of a given $F$ level (elastic collisions), as well as between the magnetic sublevels of different $F$ levels, belonging either to the same $J$ level or to different $J$ levels of the same term (quasi-elastic transitions). All these collisional transitions cause atomic level polarization to be relaxed or transferred between atomic levels, thus modifying (generally reducing) the polarization of the spectral line radiation [9]. When such collisional transitions are taken into account in full generality, the statistical equilibrium equations for a two-term atom with HFS can only be solved numerically and the redistribution matrix formalism cannot be applied. For this reason, the impact of these collisional processes on atomic polarization was neglected in the calculations shown in Fig. 2 of [1]. However, in the unmagnetized case, an analytical solution to the statistical equilibrium equations can still be found when accounting for collisional transitions between magnetic sublevels of the same $F$ level only. These transitions produce a relaxation of atomic polarization in each $F$ level of the upper term, quantified by the depolarizing rates $D^{(K)}(J, F)$. We made use of the ensuing redistribution matrix [9] to analyze the impact of such collisional depolarization on the scattering polarization pattern of the sodium doublet. For all HFS levels with $F > 0$, which can carry atomic alignment [5, 13], the $D^{(2)}$ depolarizing rates were taken to be equal to the rate of elastic collisions, being roughly two times larger than commonly used approximations. Under such assumptions, we could verify that their impact is negligible. This is not surprising considering that the line-core radiation of $D_1$ and $D_2$ is mainly emitted from the solar chromosphere, where the density of neutral hydrogen is relatively low, and that the wings are produced by coherent scattering processes which, by definition, are unaffected by elastic collisions. A previous theoretical investigation suggested that collisional transitions between the upper $J$ levels of $D_1$ and $D_2$ may have an impact on scattering polarization comparable to that of the transitions between magnetic sublevels of the same $F$ level [9]. Because we found the depolarization due to the latter transitions to be negligible, we may safely expect that, in a realistic RT modeling, the former will not have any appreciable impact on the scattering polarization pattern of the sodium doublet either.

## IV. THE RADIATIVE TRANSFER COEFFICIENTS

The RT coefficients of the Stokes-vector transfer equation [2], namely the emission coefficient in the four Stokes parameters and the elements of the propagation matrix, were calculated accounting for both line and continuum contributions. The scattering contribution to the line emission coefficient was obtained through the redistribution matrix dicussed in Sect. II. The thermal contribution to the line emission coefficient was calculated using an expression derived under the assumption that the electron-atom interaction is described by a dipole operator [14]. The line contribution to the elements of the propagation matrix was obtained from the dipole matrix elements for an atomic system with both fine structure and HFS, accounting also for the incomplete Paschen-Back effect [15]. Because the radiation field in the solar atmosphere is highly diluted, stimulated emission was safely neglected in our calculations [2].

The scattering contribution to the continuum emission coefficient was evaluated assuming coherent scattering in the observer's frame [16]. The corresponding extinction coefficient and the thermal contribution to the continuum emission coefficient were evaluated using the RH code [17]. Regarding the elements of the propagation matrix, continuum processes only contribute to the absorption coefficient for intensity (i.e., continuum opacity). Indeed, the continuum dichroism phenomena can be safely neglected in the visible range of the solar spectrum [2].

## V. THE ITERATIVE METHOD

The results presented in [1] were obtained by solving the non-equilibrium problem of the generation and transfer of polarized spectral line radiation taking into account all the physical mechanisms mentioned above. To obtain the self-consistent solution we iteratively solved the statistical equilibrium equations (implicit in the redistribution matrix) and the radiative transfer equation. At each iterative step we used the results of one set of equations as the input for the other. We applied a Jacobi-type iterative scheme [18, 19], obtained by generalizing the approach that was recently developed for the case of a two-level atom, accounting for frequency coherence in scattering processes and in the presence of an external magnetic field [20], to the case of the more complex atomic model considered in the present work. As the initial guess for the radiation field, we considered the converged solution of the RT problem for intensity only, obtained with the RH code [17]. For this calculation, an atomic model of sodium consisting of 12 levels (the ground level of neutral sodium, ten excited levels including the upper levels of $D_1$ and $D_2$, and the ground level of ionized sodium) was considered. This converged solution also supplied the population of the ground level of neutral sodium, which is kept fixed when iteratively solving the RT problem with polarization.

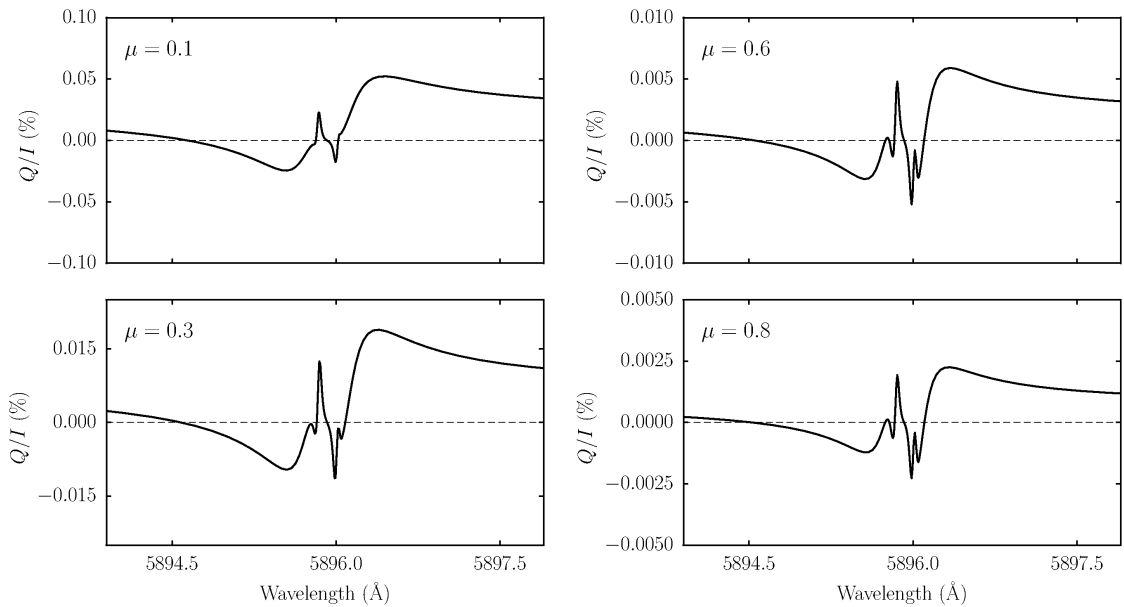

FIG. 1. Center-to-limb variation of the $D_1$ scattering polarization patterns. In these calculations a tangled volume-filling magnetic field of 15 G is considered (see text). Each panel shows the results for a different line of sight, as indicated by their respective labels. The reference direction for positive Stokes $Q$ is taken parallel to the nearest solar limb. For the considered magnetic field distribution the Stokes $U/I$ signals are zero and, therefore, are not shown here. The scale of each panel has been adjusted as the scattering polarization amplitude decreases when approaching the disk center.

## VI. THE ATMOSPHERIC MODEL

The calculations presented in this work were carried out in a semiempirical model of the quiet solar atmosphere [21]. For the relative abundance of sodium, we used $\log(N_{Na}/N_H) = 6.33$ [22]. All the collisional rates were evaluated with the RH code [17], in which the approximate formula of Unsöld [23] was used to compute the rate of elastic collisions.

## VII. CALCULATIONS IN THE PRESENCE OF TANGLED MAGNETIC FIELDS

The numerical code used to solve the RT problem with polarization allows considering not only magnetic fields with any given orientation (hereafter, deterministic fields), but also tangled magnetic fields. For the latter case, here we discuss two possible realizations: A) magnetic fields with a fixed strength whose orientation changes over scales below the mean free path of the line's photons (micro-structured fields), and B) magnetic fields with a fixed strength and whose orientation changes over scales larger than the mean free path of the line's photons, but below the resolution element of a hypothetical instrument. For the calculations we presented in [1], we considered scenario (B), taking tangled fields with an isotropic distribution over orientations. We note that, in order to reproduce canopy-like fields, we have also made numerical tests considering horizontal magnetic fields with randomly distributed azimuths, although such results are not presented here.

When considering an isotropic field distribution, scenario (A) is achieved by analytically averaging the considered redistribution matrix over such distribution of orientations [19]. At any spatial point, the net longitudinal component of the magnetic field vanishes, and consequently the MO effects that could in priniciple induce a rotation of the plane of linear polarization, quantified by the $\rho_V$ coefficient of the propagation matrix, do not operate [20]. Scenario B) is implemented by averaging the Stokes profiles of the emergent radiation obtained through multiple realizations of the RT problem, changing only the magnetic field orientation for each of them. This average, and the selection of the magnetic field orientations, is carried out following a suitable quadrature rule over the solid angle. In this case, MO effects are operative [24].

Like Fig. 2 of [1], Fig. 1 shows the results of calculations carried out considering 15 G isotropic tangled magnetic fields corresponding to scenario (B), for which the impact of MO effects on the scattering polarization wings of the sodium doublet is appreciable. For this realization, we considered eight distinct field inclinations according to a Gauss-Legendre quadrature rule (four between 0° and 90° and the other four between 90° and 180°) and eight equally spaced azimuths following the trapezoidal rule, thereby averaging over 64 realizations. When scenario (B) is implemented in our RT modeling, with magnetic field strengths of the order of gauss, the amplitude of the $D_1$ linear polarization signal around the line core remains significant. Figure 1 also shows, as expected, that the amplitude is largest for lines of sight close to the solar limb and decreases for increasing values of $\mu = \cos\theta$, where $\theta$ is the heliocentric angle. It is noteworthy that significant line-core $Q/I$ signals are still found when $\mu$ takes values relatively close to unity (i.e., when approaching the disk center), in agreement with observations [25].

## VIII. CALCULATIONS IN THE PRESENCE OF DETERMINISTIC MAGNETIC FIELDS

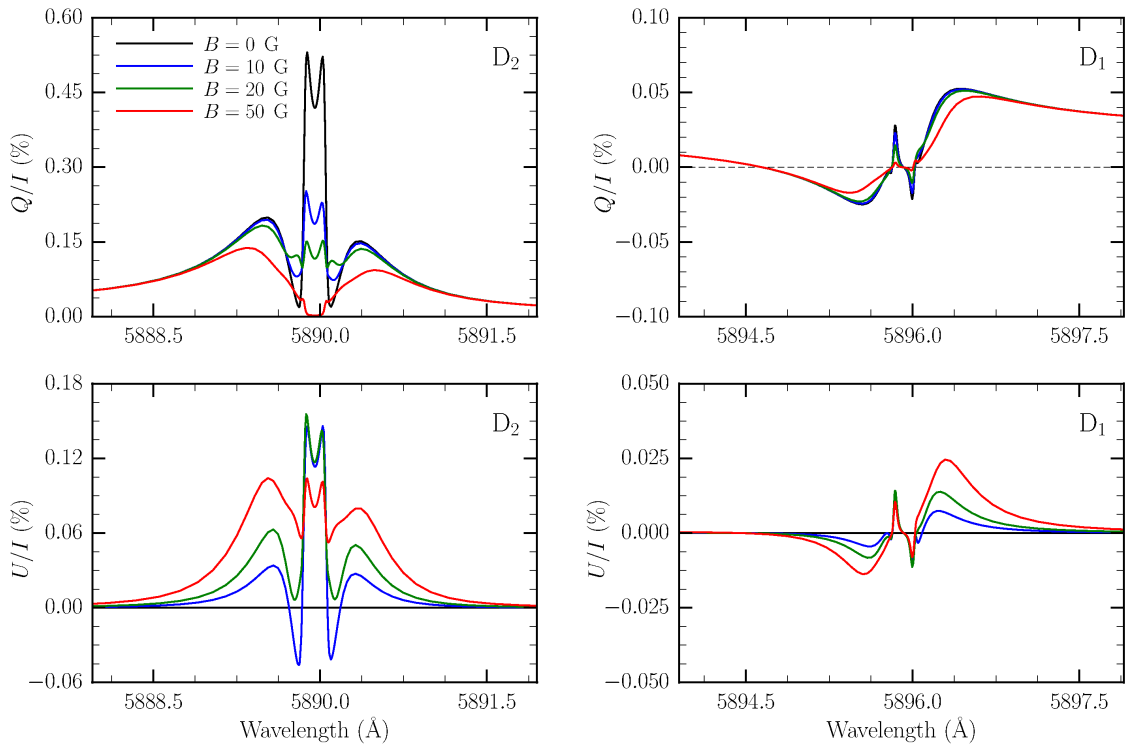

FIG. 2. Stokes $Q/I$ and $U/I$ profiles in the spectral ranges containing the $D_2$ and $D_1$ lines of Na I, in the presence of horizontal fields. The reference direction for positive Stokes $Q$ is taken parallel to the nearest solar limb. The calculations have been carried out for a line of sight close to the limb, with $\mu = 0.1$, in the presence of a horizontal magnetic field laying in the plane defined by the vertical and the line of sight, with different strengths (see legend). The magnetic sensitivity in the core of the lines is due to the Hanle effect (i.e., the magnetically induced modification of interference (a) and (b) discussed in Sect. I), with interference (a) causing the very significant impact that is appreciable in the core of the $Q/I$ and $U/I$ profiles of $D_2$. The magnetic sensitivity found in the $Q/I$ and $U/I$ wings of both lines is caused by the magneto-optical (MO) effects.

We have also performed calculations in the presence of deterministic magnetic fields, for a line of sight close to the solar limb ($\mu = 0.1$). In particular, we considered horizontal magnetic fields contained in the plane defined by the local vertical and the line of sight (see Fig. 2), and vertical magnetic fields (Fig. 3). In the former case, we highlight the magnetic sensitivity of the wings of both lines due to the MO effects, and the depolarization and rotation of the plane of linear polarization in the line cores produced by the Hanle effect. In the presence of vertical fields, for which the longitudinal component is considerably smaller, the sensitivity to MO effects is much more modest, and the magnetic modification of the interference between different $F$ levels gives rise to an enhancement of the $Q/I$ signal in the $D_2$ core [13] and a depolarization of the $D_1$ core. For this geometry, if the magnetic field strength is around 50 G, a linear polarization signal produced by the Zeeman effect begins to be appreciable in both lines.

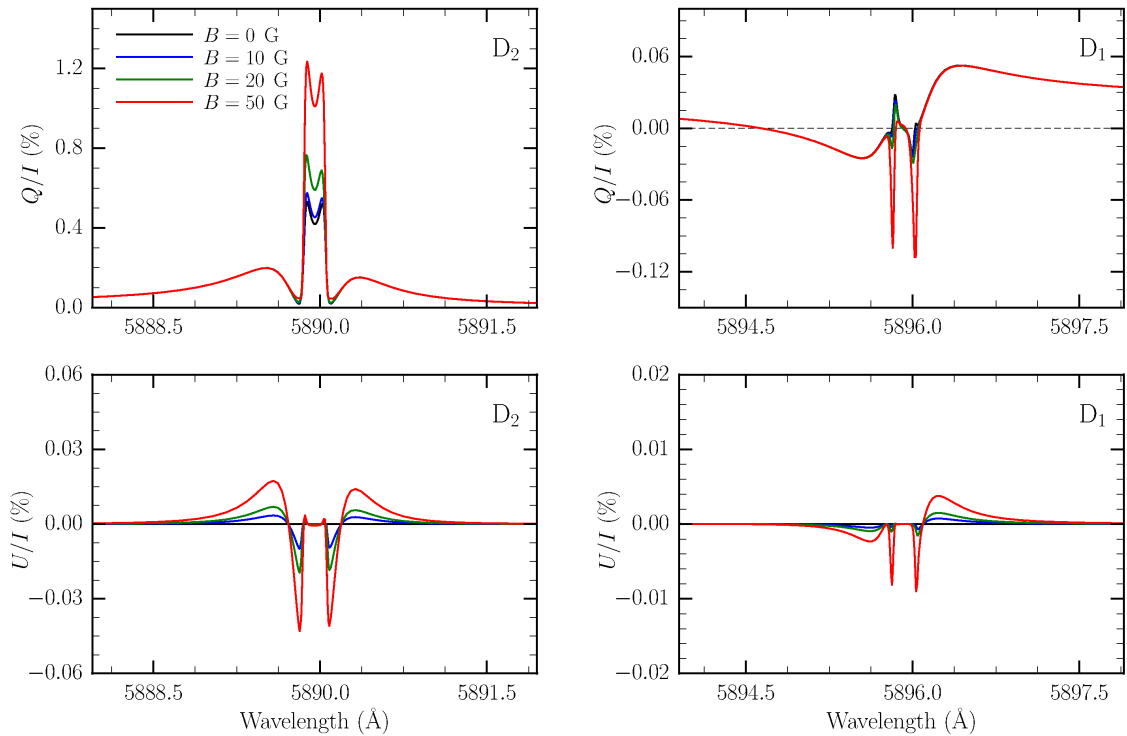

FIG. 3. Stokes $Q/I$ and $U/I$ profiles in the spectral ranges containing the $D_2$ and $D_1$ lines of Na I, in the presence of vertical fields. The strengths of the considered fields are shown in the legend. The reference direction for positive Stokes $Q$ is taken parallel to the solar limb. The calculations have been carried out for a line of sight with $\mu = 0.1$. The very clear enhancement of the $Q/I$ signal by vertical magnetic fields seen at the core of $D_2$ is due to the magnetically induced modification of the quantum interference between the upper $F$ levels of this line (i.e., to interference (b) discussed in Sect. I). The magnetic sensitivity found in the wings of both lines for vertical fields up to about 20 G is mainly due to the MO effects, while at 50 G the signatures of the familiar Zeeman effect are clearly appreciable in the $Q/I$ and $U/I$ signals.